# Bulk cyclotron resonance in the topological insulator $Bi_2Te_3$

Dmytro L. Kamenskyi [1,2], Artem V. Pronin [3], Hadj M. Benia [4], Victor P. Martovitskii [5], Kirill S. Pervakov [5], Yurii G. Selivanov [5]

[1] High Field Magnet Laboratory (HFML-EMFL) and FELIX Laboratory, Radboud University, 6525 ED Nijmegen, The Netherlands
[2] Experimentalphysik V, Center for Electronic Correlations and Magnetism, University of Augsburg, 86159 Augsburg, Germany
[3] 1. Physikalisches Institut, Universität Stuttgart, 70569 Stuttgart, Germany
[4] Centre de Développement des Technologies Avancées (CDTA), Baba Hassen, Algiers, Algeria
[5] P. N. Lebedev Physical Institute of the RAS, 119991 Moscow, Russia

**Abstract:** We investigated magneto-optical response of undoped $Bi_2Te_3$ films in the terahertz frequency range (0.3 – 5.1 THz, 10 – 170 cm$^{-1}$) in magnetic fields up to 10 T. The optical transmission, measured in the Faraday geometry, is dominated by a broad Lorentzian-shaped mode, whose central frequency linearly increases with applied field. In zero field, the Lorentzian is centered at zero frequency, representing hence the free-carrier Drude response. We interpret the mode as a cyclotron resonance (CR) of free carriers in $Bi_2Te_3$. Because the mode's frequency position follows a linear magnetic-field dependence and because undoped $Bi_2Te_3$ is known to possess an appreciable number of bulk carriers, we associate the mode with a *bulk* CR. In addition, the cyclotron mass obtained from our measurements fits well the literature data on the bulk effective mass in $Bi_2Te_3$. Interestingly, the width of the CR mode demonstrates a behavior non-monotonous in field. We propose that the CR width is defined by two competing factors: impurity scattering, which rate decreases in increasing field, and electron-phonon scattering, which rate exhibits the opposite behavior.

---

## 1. Introduction

From the theory point of view, a three-dimensional (3D) topological insulator (TI) possesses insulating bulk and conducting surfaces, the conduction channels at surfaces being spin-polarized [1–4]. Since the spin polarization can potentially be utilized in spintronic devices, topological insulators have attracted a lot of attention in the past years [5,6]. In practice, the real samples of 3D topological insulators often conduct not only on their surfaces, but also in the bulk. Considerable efforts have been made to understand and separate the properties of surface and bulk charge carriers. These properties can particularly be studied via different spectroscopic techniques, such as angle-resolved photoemission spectroscopy (ARPES) or optical and magneto-optical spectroscopy. The optical conductivity and cyclotron resonance (CR) of a number of 3D TI materials have been reported in the literature. Perhaps the most studied family of such TIs is the bismuth selenide – bismuth telluride series, $Bi_2(Te_{1-x}Se_x)_3$, which also includes the undoped members, $Bi_2Te_3$ and $Bi_2Se_3$ [7-9]. In this study, we concentrate on $Bi_2Te_3$. Namely, we investigate experimentally the CR in this compound. Surprisingly, the published CR measurements performed on this well-studied TI produce rather diverging results [10-13] with the absorption features being generally of rather complex shapes. One of the reasons for such diversity might be the sample-dependent variation between the surface and bulk contributions, which, in turn, greatly depend on the exact position of the Fermi level.

Unlike in the majority of previous reports [11-13], the CR absorption observed in our study can be well described by a *single* Lorentzian-shaped mode (which is rather consistent with the earliest study on this issue from 1999 [10]). We believe the absorption we detect is of bulk origin. Thus, our findings might be useful for the proper interpretations of the CR modes in doped $Bi_2Te_3$, where the balance between the surface and bulk-states contributions can be shifted towards the former, but the bulk still cannot be completely ignored.

## 2. Materials and Methods

We grew thin layers of $Bi_2Te_3$ on (111)-oriented $BaF_2$ substrates by molecular beam epitaxy [14]. For the growth, we used binary $Bi_2Te_3$ and elemental Te. This is different from the standard practice, when elemental (Bi and Te) sources are utilized with the typical flux ratio of Te/Bi being about 10 to 20. Using $Bi_2Te_3$ and Te allowed us to reduce this ratio to the values below 1 and, hence, to precisely control the stoichiometry of the growing layer. Along with the employed "ramp up" growth procedure [15], these two approaches successfully suppress twin formation in the growing films. X-ray diffraction (XRD) ϕ scans about the [0 0 1] axis on the asymmetric (1 0 10) reflection revealed only 120°-periodic peaks and confirmed that the films obtained by this method are either single-domain or have a very small twin volume fraction (3-7 % for the films with 1 cm² area) with the *c* axis of $Bi_2Te_3$ being perpendicular to the substrate surface. To the best of our knowledge, this thin-film growth method is unique.

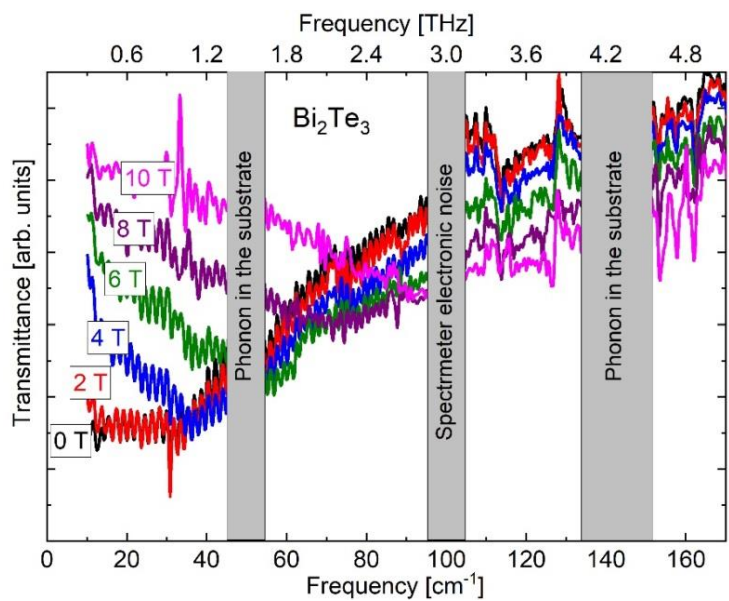

**Figure 1.** Raw transmission spectra of $Bi_2Te_3$ films on $BaF_2$ substrates as obtained in magnetic fields of up to 10 T. The areas with low signal due to either the substrate phonons or spectrometer electronic noise are shaded. The signal-to-noise ratio is best at around 80 cm$^{-1}$ and becomes appreciably lower as frequency increases (see also Figure 2), preventing thus any meaningful measurements of the CR mode at the fields higher than 10 T.

In order to prevent possible influence of atmospheric oxygen and water, we have developed a method to cover the TI films *in situ* with optically friendly protecting layers of $BaF_2$ [16]. We have found that 30 – 50 nm of $BaF_2$ provide the optimal protection. Our measurements have shown that the $BaF_2$ cap layers affect neither crystal-structure parameters nor optical properties at the frequencies of interest.

The sample used in this study was thoroughly characterized by XRD, scanning electron microscopy (SEM), atomic force microscopy (AFM), and ARPES. The results of these investigations, presented in the Supplemental Material, confirm high structural and morphological quality of the film and show that the film possesses the topological surface electronic states as well as the states in the bulk conduction band.

For optical measurements, we utilized the infrared optical setup available at the High Field Magnet Laboratory in Nijmegen [17]. This setup consists of a commercial Fourier-transform infrared (FTIR) spectrometer (Bruker IFS113v) combined with a continuous-field 33-Tesla Bitter magnet. A detailed description of this setup could be found elsewhere [18]. The measurements were performed in the Faraday geometry [19] at 2 K. A mercury lamp was used as a radiation source. The far-infrared radiation was detected using a custom-made silicon bolometer operating at 1.4 K. The FTIR spectra were recorded in a number of magnetic fields from 0 to 30 T. The optical data were collected between 10 and 170 cm$^{-1}$ (300 – 5100 GHz), using a 200-$\mu$m Mylar beamsplitter and a scanning velocity of 50 kHz. At each field, at least 100 scans were averaged. As will be seen below, the data obtained in the fields above 10 T cannot be used in our analysis because of a low signal-to-noise ratio. Thus, in this study we concentrate on the measurements performed between 0 and 10 T.

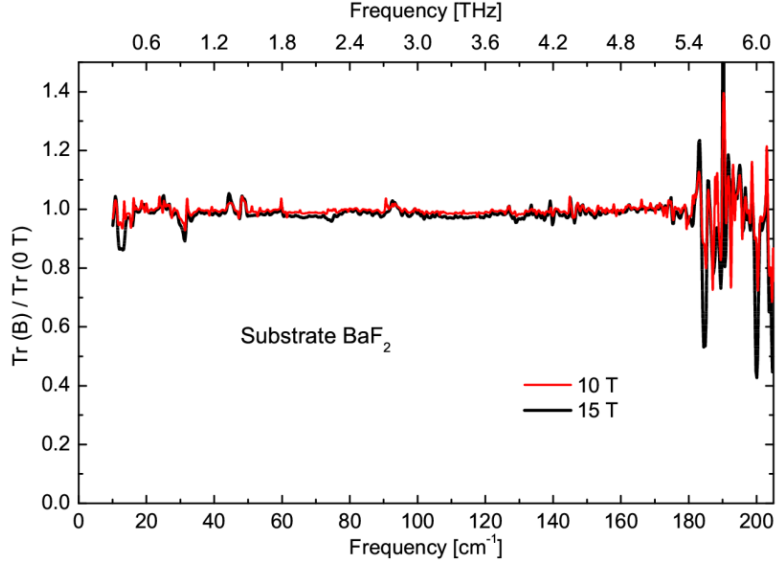

**Figure 2.** Frequency-dependent transmission of a bare BaF$_2$ substrate at 10 and 15 T normalized to its zero-field spectrum. The noise at high frequencies is due the experimental setup. The Figure is meant to demonstrate: (i) the absence of any field-induced changes in the optical spectra of BaF$_2$ and (ii) the frequency limits of the setup used.

## 3. Results and Discussion

In Figure 1, we show raw transmission data measured though the 115-nm-thick Bi$_2$Te$_3$ film on a 0.49-mm-thick BaF$_2$ substrate (cf. Figure S2 in the Supplemental Material) in magnetic fields *B* up to 10 T. We note that all the measurements reported in this study are performed on a single sample. As seen from Figure 2, the substrate has no detectable field dependence. Hence, all the field-induced changes come from the film. We note that the BaF$_2$ substrate has intense phonon modes at roughly 50 and 140 cm$^{-1}$ [20]. Thus, accurate measurements around these frequencies are impossible. The spectra of Figure 1 are dominated by a single broad mode, which position shifts to higher frequencies in increasing field. The spectra can be fitted by a single Lorentzian, as exemplified in Figure 3 for 2, 4, and 6 T. In zero field, the Lorentzian central frequency is zero, i.e., the observed absorption mode is due to free carriers (Drude conductivity). The field evolution of the mode can be traced in Figure 1: with increasing field, the mode shifts upwards and eventually goes above 100 cm$^{-1}$, i.e., in the range, where the signal-to-noise ratio is worsened by the spectrometer noise and the phonons in the substrate (this prevents a meaningful spectra analysis in higher fields). Still, the shift of the mode in the applied magnetic field is apparent and can straightforwardly be interpreted as a magnetic-field-induced free-carrier localization or, in other words, a cyclotron-resonance absorption.

The Lorentzian-fit results for this CR absorption mode in the fields from 0 to 10 T are shown in in Figure 4. One can see that the central frequency of the absorption line is linear in field (left panel). This immediately signals that the electronic band(s) responsible for the observed absorption have a quadratic dispersion relation. For linear electronic bands, the field dependence of the CR lines is supposed to have a square-root dependence on applied field [21]. Thus, following the Occam's razor principle, we conclude that the mode is due to bulk (i.e., not linear, not topological) electronic bands. This conclusion is in full agreement, e.g., with ARPES [9] and quantum-oscillations [22] measurements, which show that the Fermi level in undoped Bi$_2$Te$_3$ crosses the bulk conduction band and hence there exists a large bulk Fermi surface.

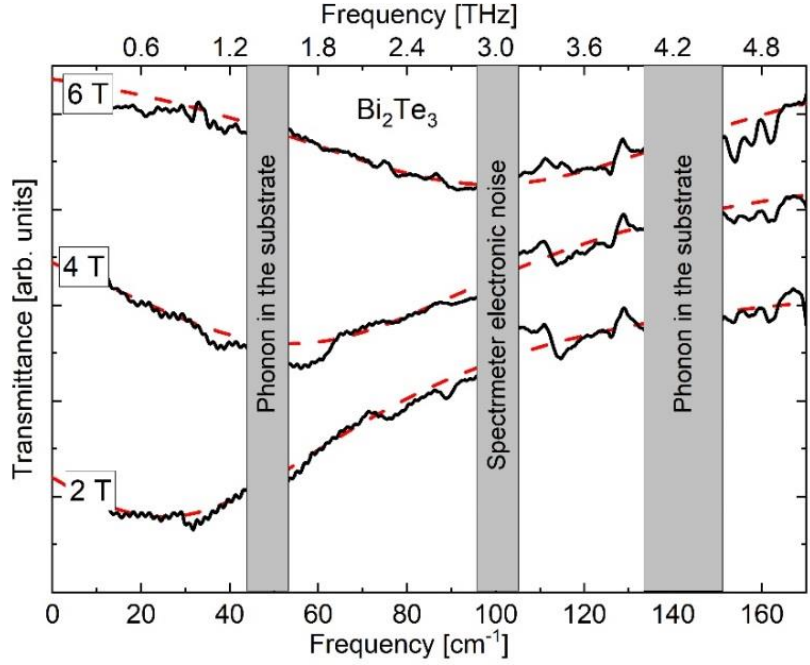

**Figure 3.** Examples of Lorentzian fits of the transmission spectra from Figure 1 for a few magnetic-field strengths as indicated. The raw experimental data are smoothed, using a Savitzki-Golay method [23]. Note that the spectra for 4 and 6 T are shifted upwards for clarity.

We note that weak modes due to the surface conduction channels may exist on top of the dominating bulk absorption, but within our accuracy they cannot be resolved.

The linear field dependence of the central CR frequency, $\omega_0$, can be fitted with the standard parabolic-band expression, connecting the slope of $\omega_0(B)$ and the carrier cyclotron mass, $m^*$: $\omega_0 = eB/m^*c$ (CGS units are used, $e$ is the elementary charge, $c$ is the speed of light). This fit is shown in the left panel of Figure 4 as a straight line and provides $m^* = 0.1 m_e$ ($m_e$ is the free-electron mass). This value is in very good agreement with the available literature data on the bulk effective mass in $Bi_2Te_3$ [24]: $m^* = 0.109 m_e$ for the response perpendicular to the $c$-axis, which we do probe in our transmission experiment with unpolarized light. This match provides another confirmation for the correctness of our interpretation. We would like to note here, that the complete agreement between the calculated electronic band structure of $Bi_2Te_3$ and the entire body of the available experimental work is still to be achieved, as emphasized in a recent review [25].

Finally, we turn to the width of the absorption band. As one can see from the right panel of Figure 4, the full width at half maximum (FWHM) of the band demonstrates a non-monotonous field dependence: in low fields, it decreases with increasing $B$ and then, starting at approximately 5 T, the FWHM starts growing with the applied field. The initial decrease of FWHM can be naturally explained by the decreasing cyclotron-orbit radius with increasing $B$ and the consequent decrease of impurity scattering. The reason for the CR mode broadening in $B > 5$ T is not entirely clear. We propose that this could be due to the increased electron-phonon scattering. In higher fields, the CR mode approaches the frequencies, where phonon density grows (roughly, above 40 cm$^{-1}$; cf. the left panel of Figure 4 and Ref. [26], where the phonon density for $Bi_2Te_3$ was calculated) and hence the rate of the electron-phonon scattering starts to increase, leading to the observed total broadening of the CR line according to the Matthiessen rule.

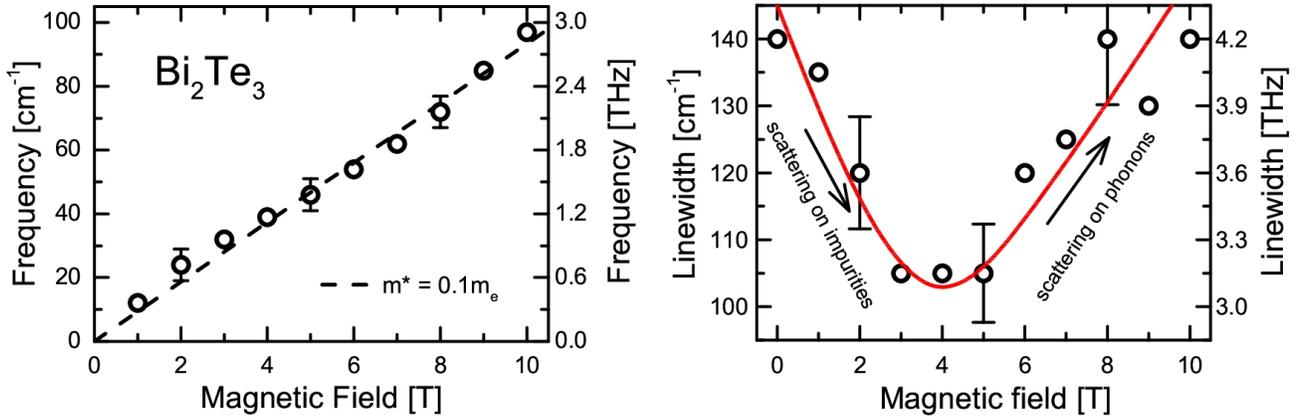

**Figure 4.** CR-line central frequency (left frame) and FWHM (right frame) versus applied magnetic field. The dashed line is a fit with $m^*=0.1m_e$. The red solid line is a guide for the eye.

## 5. Conclusions

We have investigated the magneto-optical response of undoped $Bi_2Te_3$ films at terahertz frequencies and in magnetic fields of up to 10 T. We observed an intense CR line, which can be fitted with a single Lorentz oscillator. The central frequency of the CR increases linearly with applied field, signaling the bulk origin of this resonance. In addition, we found the "in-plane" cyclotron mass, $m^*=0.1m_e$, which matches well the literature data for bulk $Bi_2Te_3$. The width of the CR mode demonstrates a behavior non-monotonous in field. We propose that the CR width is defined by two competing factors: impurity scattering, which rate decreases in increasing field, and electron-phonon scattering, which rate demonstrates the opposite behavior. We believe our findings can be exploited in future measurements of the surface-states CR in $Bi_2Te_3$ to disentangle the bulk and surface contributions.

**Acknowledgements:** This work was partially supported by the PRIME program of the German Academic Exchange Service (DAAD) with funds from the German Federal Ministry of Education and Research (BMBF) and by the Russian Foundation for Basic Research (grant No. 20-02-00989). H.M.B. acknowledges funding from the German Research Foundation (DFG) via Project No. BE 5190/1-1. We also acknowledge the support of HFML-RU/NWO, member of the European Magnetic Field Laboratory (EMFL).

# Bulk cyclotron resonance in the topological insulator $Bi_2Te_3$

## Supplemental material

D. L. Kamenskyi [1,2], A. V. Pronin [3], H. M. Benia [4], V. P. Martovitskii [5], K. S. Pervakov [5], Yu. G. Selivanov [5]

[1] High Field Magnet Laboratory (HFML-EMFL) and FELIX Laboratory, Radboud University, 6525 ED Nijmegen, The Netherlands
[2] Experimentalphysik V, Center for Electronic Correlations and Magnetism, University of Augsburg, 86159 Augsburg, Germany
[3] 1. Physikalisches Institut, Universität Stuttgart, 70569 Stuttgart, Germany
[4] Centre de Développement des Technologies Avancées (CDTA), Baba Hassen, Algiers, Algeria
[5] P. N. Lebedev Physical Institute of the RAS, 119991 Moscow, Russia

The film used in the magneto-optical measurements was characterized by a number of experimental probes as described below.

## A. Structural and morphological characterization

For X-ray diffraction (XRD) measurements we used a Panalytical MRD Extended diffractometer. In Figure S1, the results of these studies are shown. In panel (a), a series of (0 0 l) reflections from the film is seen together with the intensive (hhh) peaks from the $BaF_2$ substrate. Thus, the growth of a highly oriented single-phase layer with the basal plane (0 0 1) parallel to the $BaF_2$ substrate (1 1 1) cleavage plane is evidenced. The high crystalline quality is supported by the absence of any noticeable broadening when going from the (0 0 3) to the (0 0 36) reflection peaks. As one can see from panel (b), the full width at the half maximum of the (0 0 15) rocking curve is $\Delta\omega=0.082°$. The somewhat asymmetric shape of the curve indicates the presence of the anti-site defects, responsible for impurity scattering.

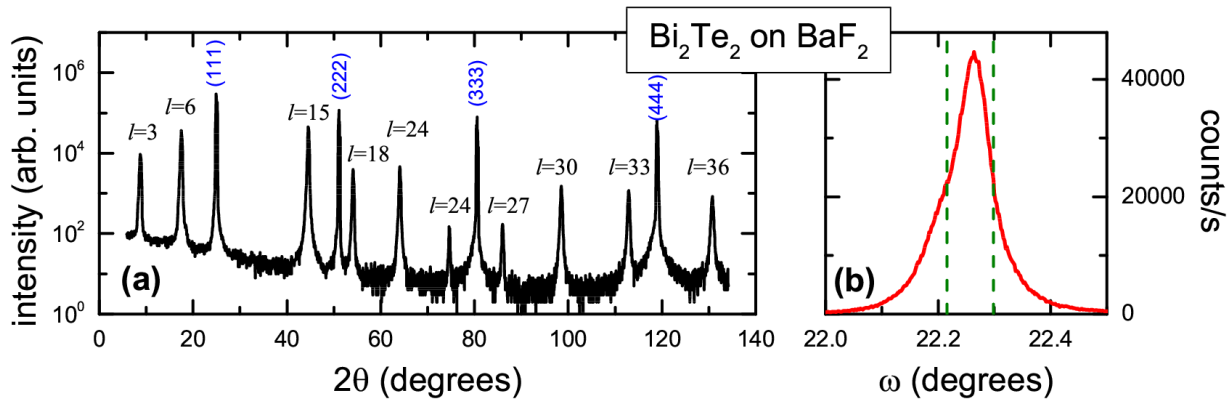

**Figure S1.** Panel (a): X-ray diffraction scan of the $Bi_2Te_3$ epitaxial film used in this study. A series of the (hhh) peaks belongs to the substrate, while all the (00l) reflections are from the $Bi_2Te_3$ layer. Panel (b): Rocking curve for the (0 0 15) reflection peak.

We used a JSM-7001F scanning electron microscope (SEM) to obtain the cleaved cross-section images of the film. One of such images is shown in Figure S2 (panel (a)). From this picture, the thicknesses of the $Bi_2Te_3$ film and of $BaF_2$ cap layer were evaluated to be 115 and 49 nm, respectively.

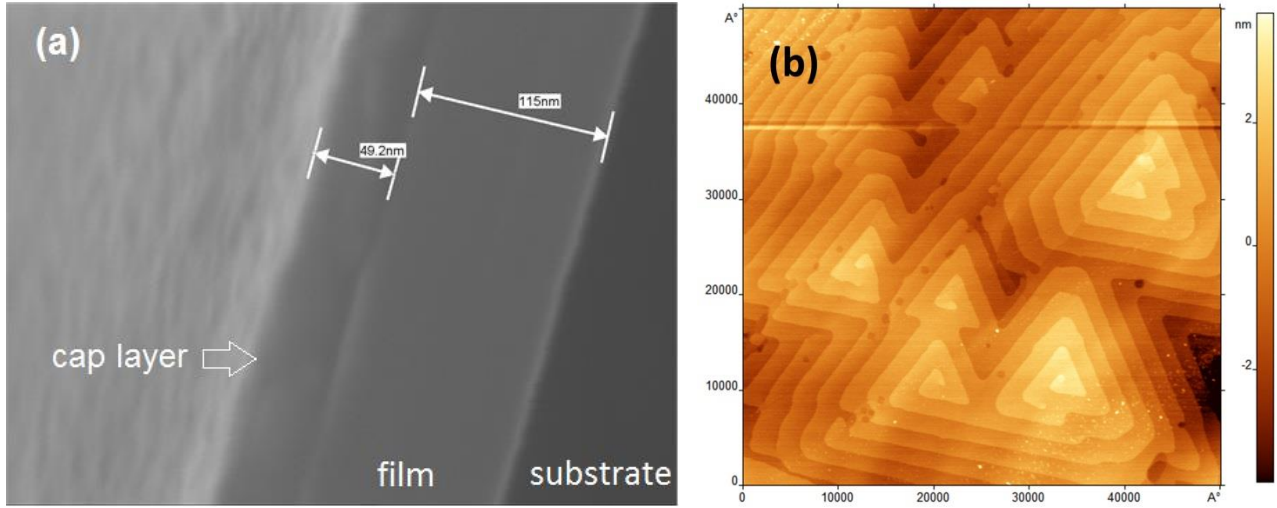

**Figure S2.** Panel (a): Cross-section SEM image of the $Bi_2Te_3$ film with a $BaF_2$ cap layer obtained with the 300000-times magnification. Panel (b): Surface characterization of the $Bi_2Te_3$ film by AFM. The scanned area is 50 000 Å × 50 000 Å. The vertical scale is color-coded as shown in the bar on the right-hand side.

The morphology of the studied film was explored by means of the atomic force microscopy (AFM) in the tapping mode using the NT-MDT Solver 47 Pro system. For these measurements, the barium fluoride capping layer was removed by the procedure described in Ref. [S1]. A representative (5 × 5 μm) AFM scan of the investigated $Bi_2Te_3$ film is shown in panel (b) of Figure S2. Regular triangular pyramids with large domain terraces and 1 nm high steps indicate high crystal quality. Observed spiral-like growth was reported previously [S2, S3] and is believed to promote formation of twin-free films. All triangular domains are oriented in the same direction evidencing single domain sample with trigonal symmetry.

**B. Photoemission characterization**

The ARPES measurements were performed with a hemispherical SPECS HSA3500 electron analyzer, characterized by an energy resolution of about 10 meV. Monochromatized He I (21.2 eV) radiation was used as photon source. During the measurements, the sample was cooled with the aid of liquid nitrogen to 100 K. Prior the measurements, the surface of the samples was cleaned by several sputter-anneal cycles (Argon sputtering: 500 eV/30 min; Annealing: 260 ℃/15 min). The results of the measurements are shown in Fig. S3.

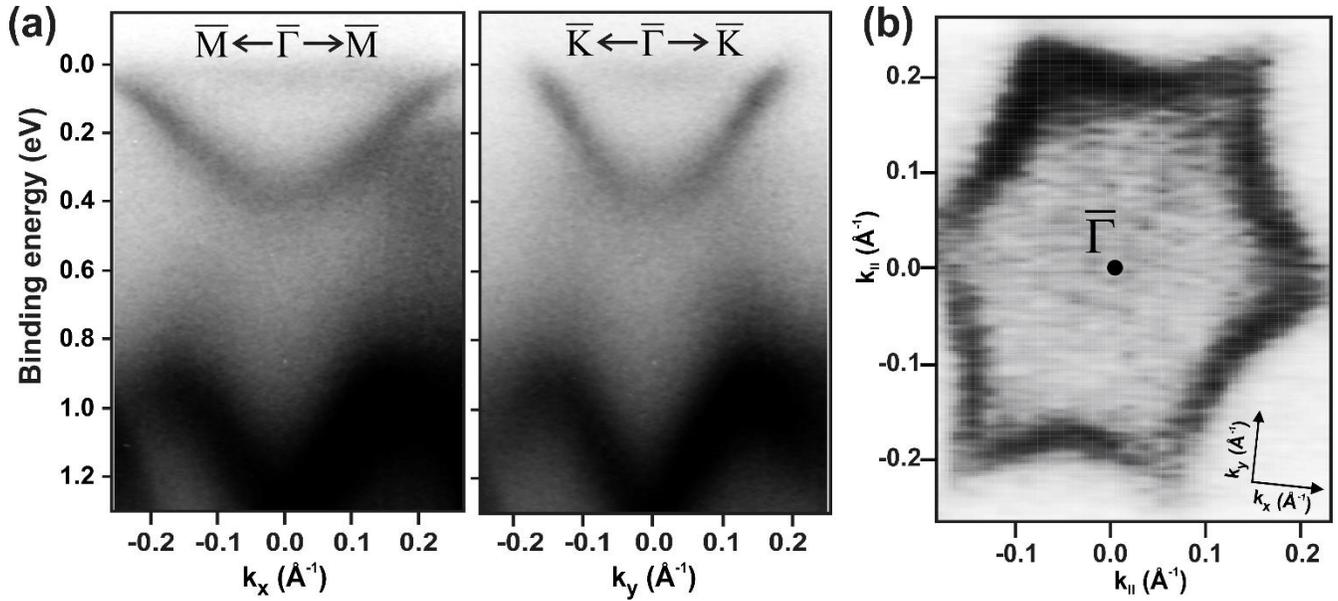

**Figure S3.** (a) Experimental band structure of $Bi_2Te_3$ film along $\bar{M}$ - $\bar{\Gamma}$ - $\bar{M}$ and $\bar{K}$ - $\bar{\Gamma}$ - $\bar{K}$ directions of the Brillouin zone, and, (b) the corresponding Fermi surface recorded at 100 K. The surface state cone clearly appears in the measurements. Very near the Γ point, the linearity of the surface bands is not perfect due to an admixture of the bulk states (cf. the ARPES data from Ref. [S4]). Note that the surface states provide a much sharper ARPES images, than the bulk states. The bulk states near the Fermi level around the Γ point could be identified in Ref. [S4], where photon energy of 48 eV was employed. At this high energy, the cross-sections for both, surface and bulk, bands are strong. We believe these very bulk states are responsible for the observed cyclotron resonance.